\documentclass{elsart1p}
\usepackage{graphicx}
\usepackage{amssymb}
\usepackage{epsfig} 
\def\beq{\begin{equation}}
\def\eeq{\end{equation}}
\def\beqa{\begin{eqnarray}}
\def\eeqa{\end{eqnarray}}
\begin{document}
\begin{frontmatter}
\title{Heavy-Flavor Production at Accelerators}
\author{Nikolaos Kidonakis\thanksref{label1}}
\thanks[label1]{This work was supported by the 
National Science Foundation under Grant No. PHY 0555372.}
\address{Kennesaw State University, Kennesaw, GA 30144, USA}
\ead{nkidonak@kennesaw.edu}
\ead[url]{http://science.kennesaw.edu/$\sim$nkidonak/}
\begin{abstract}
I discuss heavy flavor production at hadronic facilities. I present total 
cross sections and differential distributions for top quark, bottom quark, 
and charm quark production at the Tevatron and LHC colliders, and at 
fixed-target experiments such as HERA-B. 
The calculations include complete next-to-leading order corrections as well 
as higher-order soft-gluon corrections which are important near kinematical 
threshold.
\end{abstract}
\end{frontmatter}
\section{Introduction}

QCD corrections are typically large for heavy quark production. 
The current state-of-the-art theoretical calculations of cross sections 
include higher-order corrections beyond next-to-leading order (NLO).
Soft-gluon corrections of the form $[\ln^k(s_4/m^2)/s_4]_+$,   
with $s_4$ the kinematical distance from threshold, 
are dominant near threshold and can be resummed.
At next-to-leading-logarithm (NLL) accuracy this requires one-loop 
calculations in the eikonal approximation; at next-to-next-to-leading-logarithm
(NNLL) it requires two-loop calculations.
Approximate next-to-next-to-leading order (NNLO) cross sections can be 
derived from the expansion of the resummed cross section.

In the next section I present some recent calculations for top quark pair 
production at the Tevatron and the LHC. 
This is followed by a section on single top quark production, 
and a section on bottom and charm quark production at fixed-target facilities, 
including HERA-B. 
In the last section I discuss recent work on two-loop calculations 
in the eikonal approximation that are relevant for increasing the accuracy 
of resummation for heavy quark production. 

\section{Top quark pair production}

The dominant process for top quark production in hadron colliders is pair 
production via the processes, at leading order, 
$q{\bar q}\rightarrow t{\bar t}$ and $gg \rightarrow t{\bar t}$. 
There is very good agreement of theory, including soft-gluon corrections \cite{NKRV08},
with Tevatron data \cite{CDF,D0}. The same is expected to hold with 
future LHC data.
\begin{figure}
\begin{center}
\includegraphics[width=6cm]{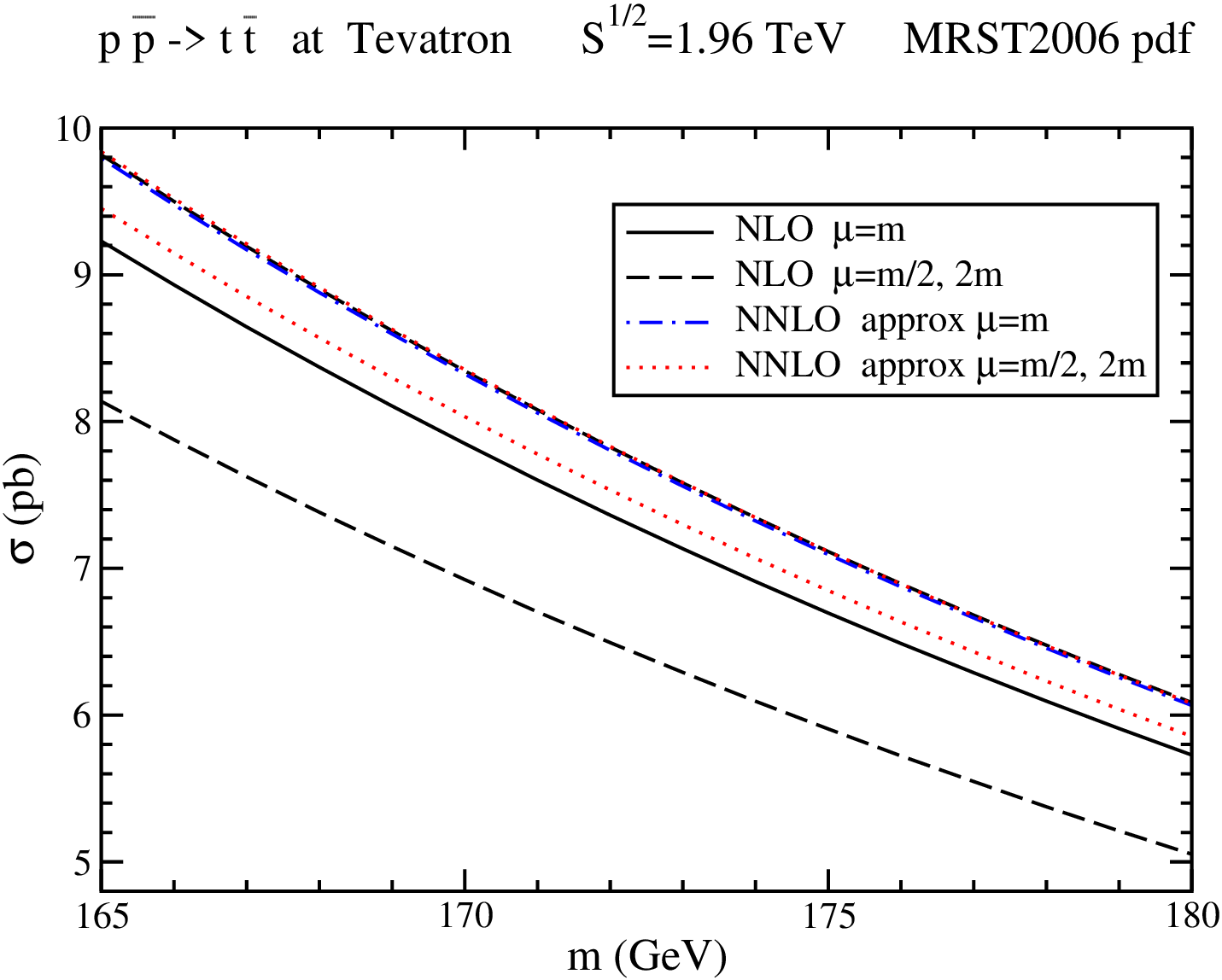}
\hspace{8mm}
\includegraphics[width=6cm]{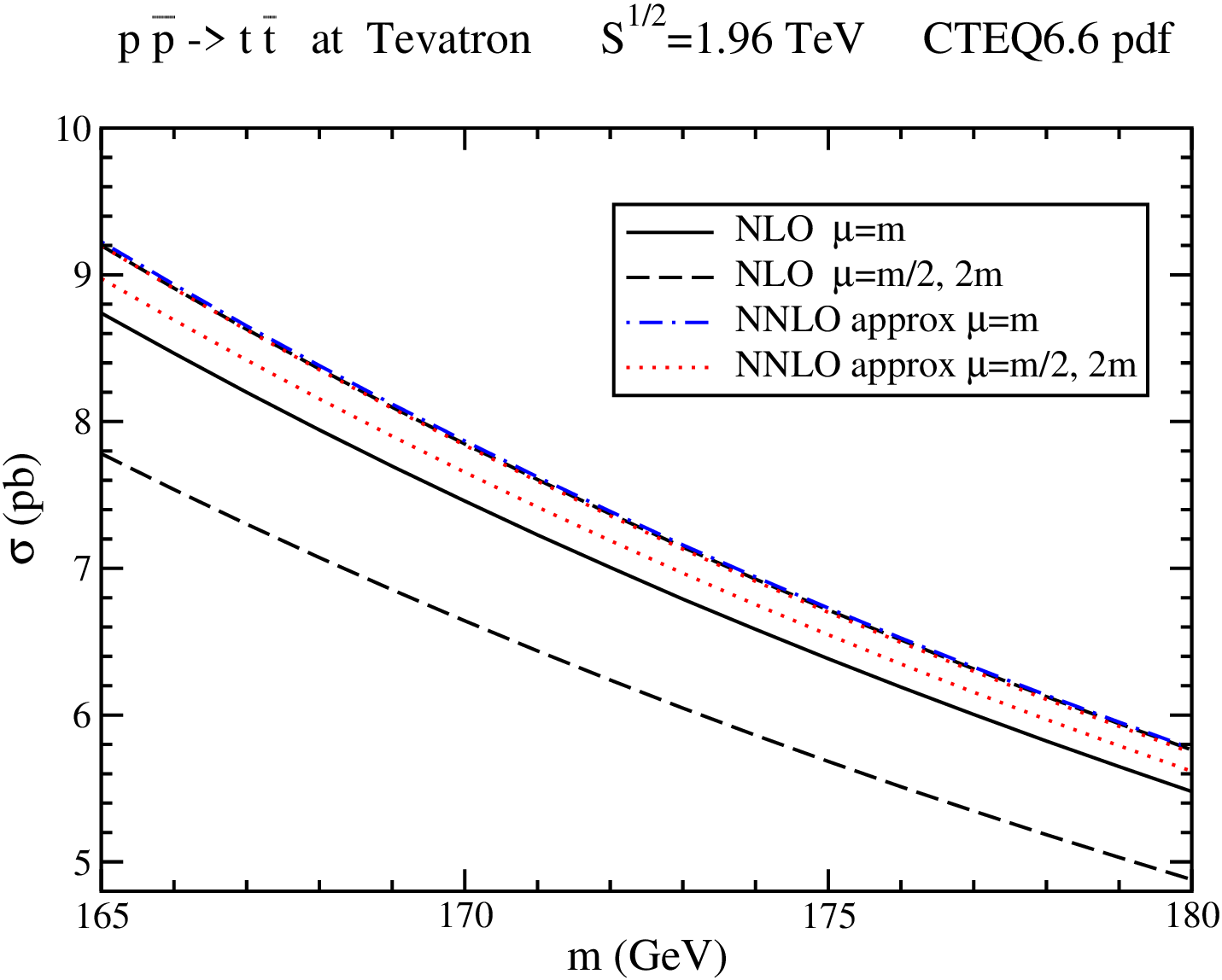}
\caption{The NLO and approximate NNLO $t {\bar t}$ cross sections in $p \overline p$
collisions at the Tevatron 
using the MRST 2006 NNLO (left) and CTEQ6.6M (right) pdf.}
\end{center}
\end{figure}

Figure 1 shows the cross section from an approximate NNLO calculation 
\cite{NKRV08} for Tevatron energy, using two different parton distribution 
function (pdf) sets, MRST 2006 NNLO \cite{MRST2006} and CTEQ6.6M \cite{CTEQ66}.

For reference, below is the cross section for a top mass of 172 GeV,
\beqa
\hspace{-7mm}
\sigma^{p{\bar p} \rightarrow t \bar t}(1.96\, {\rm TeV},
m_t=172 \,{\rm GeV}, {\rm MRST})=7.80 \pm 0.31 ^{+0.03}_{-0.27}
{}^{+0.23}_{-0.19}\; {\rm pb}=7.80 {}^{+0.39}_{-0.45} \; {\rm pb} \, ,
\nonumber
\eeqa
\beqa
\hspace{-7mm}
\sigma^{p{\bar p} \rightarrow t \bar t}(1.96\, {\rm TeV},
m_t=172 \,
{\rm GeV}, {\rm CTEQ})=7.39 \pm 0.30 ^{-0.03}_{-0.20} {}^{+0.48}_{-0.37}
\; {\rm pb}=7.39{}^{+0.57}_{-0.52} \; {\rm pb} \, .
\nonumber
\eeqa
The first uncertainty indicated is a kinematics uncertainty arising from the choice of using 
single-particle-inclusive (1PI) or pair-invariant mass (PIM) kinematics; the second uncertainty 
is from scale variation over a range $m/2 \le \mu \le 2m$, and the third uncertainty is from the pdf and varies
quite a bit between the MRST and CTEQ sets used. The total uncertainty indicated comes from the addition 
in quadrature of the three individual uncertainties.
It is interesting to note that at present the experimental and theoretical uncertainties are of similar size. It is also clear from Figure 1 that the scale dependence is 
greatly decreased at NNLO.
\begin{figure}
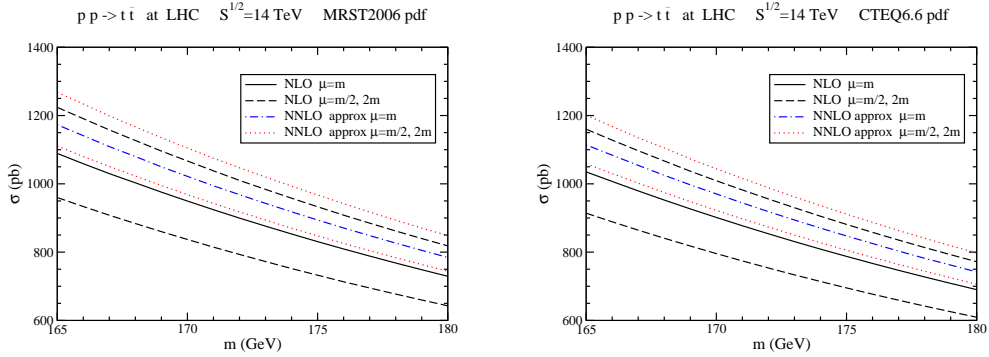

\begin{center}
\includegraphics[width=6cm]{toplhcmrstplot.eps}
\hspace{8mm}
\includegraphics[width=6cm]{toplhccteqplot.eps}
\caption{The NLO and approximate NNLO $t {\bar t}$ cross sections in $pp$
collisions at the LHC using the MRST 2006 NNLO (left) and the CTEQ6.6M (right) 
pdf.}
\end{center}
\end{figure}

Figure 2 shows the corresponding results \cite{NKRV08} for LHC energy. 
The cross sections and uncertainties 
for $m_t=172$ GeV are
\beqa
\sigma^{pp \rightarrow t \bar t}(14\, {\rm TeV},m_t=172
\,{\rm GeV}, {\rm MRST})=968 \pm  4 {}^{+79}_{-50} {}^{+12}_{-13}
\; {\rm pb}=968^{+80}_{-52} \; {\rm pb} \, ,
\nonumber
\eeqa
\beqa
\sigma^{pp \rightarrow t \bar t}(14\, {\rm TeV},m_t=172
\,{\rm GeV}, {\rm CTEQ})=919 \pm  4 {}^{+70}_{-45} {}^{+29}_{-31}   
\; {\rm pb}=919{}^{+76}_{-55} \; {\rm pb} \, .
\nonumber
\eeqa
\begin{figure}
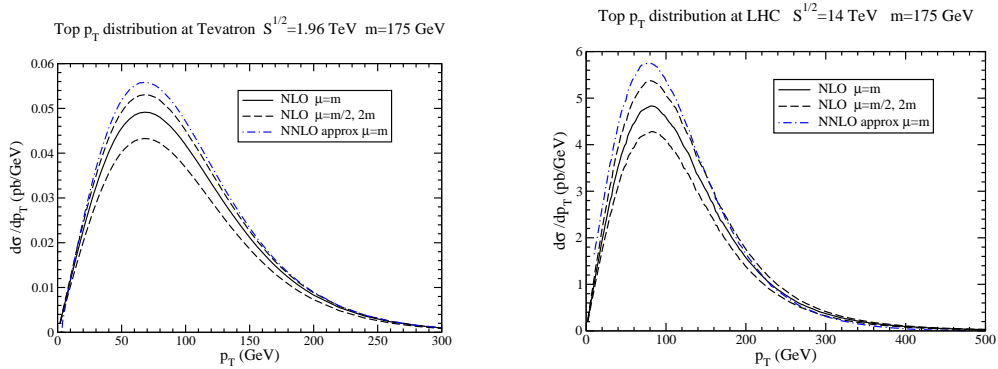

\begin{center}
\includegraphics[width=5.9cm]{pt1960plotnew.eps}
\hspace{11mm}
\includegraphics[width=5.9cm]{ptlhcplotnew.eps}
\caption{The NLO and approximate NNLO top quark $p_T$ distributions 
at the Tevatron (left) and the LHC (right).}
\end{center}
\end{figure}
Figure 3 shows the top quark transverse momentum, $p_T$, distribution 
at the Tevatron and the LHC for a top quark mass of 175 GeV \cite{NKRV03}.

\section{Single top quark production}

There is now evidence for single top production at the Tevatron \cite{CDFD0st} 
with a cross section consistent with theory \cite{NKsttev}.
The partonic processes at leading order are
the $t$ channel, $qb \rightarrow q' t$ and ${\bar q} b \rightarrow {\bar q}' t$;
the $s$ channel: $q{\bar q}' \rightarrow {\bar b} t$; and 
associated $tW$ production, $bg \rightarrow tW^-$.
In the results below we include soft-gluon corrections at NNLO and NNNLO 
\cite{NKsttev,NKstlhc,NKAPP} and use the MRST 2004 NNLO pdf \cite{MRST2004}. 
\begin{figure}
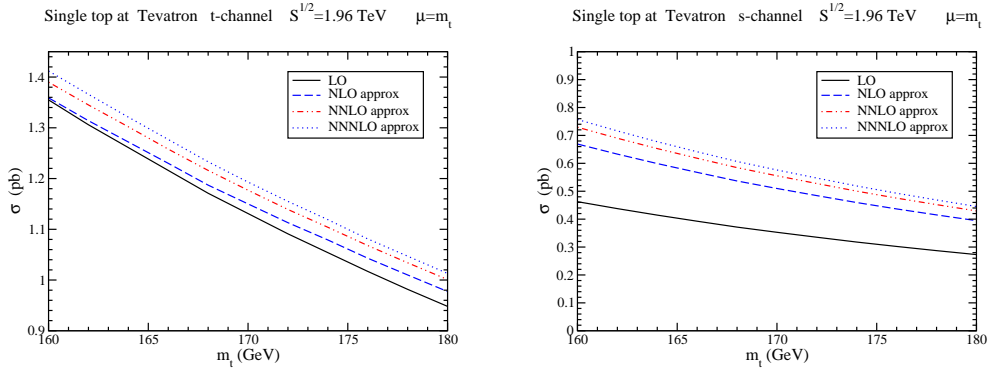

\begin{center}
\includegraphics[width=6cm]{tchtevmtplot.eps}
\hspace{8mm}
\includegraphics[width=6cm]{schtevmtplot.eps}
\caption{The single top quark cross sections at the Tevatron 
in the $t$ channel (left) and $s$ channel (right).}
\end{center}
\end{figure}

We begin with results for the Tevatron \cite{NKsttev,NKAPP}.
For the $t$ channel we find 
\beqa
\sigma^{t-{\rm channel}}_{\rm top}(1.96 \, {\rm TeV}, 
m_t=172 \,{\rm GeV})=1.14^{+0.02}_{-0.01}
\pm 0.06 \; {\rm pb}=1.14 \pm 0.06 \; {\rm pb} \, ,
\nonumber
\eeqa
where the first uncertainty is from scale variation over 
$m/2 \le \mu \le 2m$ and the second is 
from the pdf, and we add the two in quadrature for the total uncertainty.
The cross section for single antitop production is identical.

For the $s$ channel we find 
\beqa
\sigma^{s-{\rm channel}}_{\rm top}(1.96 \, {\rm TeV}, 
m_t=172 \, {\rm GeV})=0.53 \pm 0.02
\pm 0.01 \; {\rm pb}=0.53 \pm 0.02 \; {\rm pb} \, .
\nonumber
\eeqa
Again, the cross section for single antitop production is identical.

In Figure 4 we plot the cross sections for the $t$ and $s$ channels 
at various orders versus the top quark mass, with the scale set at $\mu=m_t$.

The cross section for single top production at the Tevatron via 
the $tW$ channel is smaller:
$\sigma^{tW}(1.96 \, {\rm TeV}, m_t=172 \, {\rm GeV})=0.14 \pm 0.02 
\pm 0.02 \; {\rm pb}=0.14 \pm 0.03 \; {\rm pb}$, 
with an identical result for ${\bar t} W$ production.

We continue with single top production at the LHC \cite{NKstlhc,NKAPP}. 
We note that here in the $t$ and $s$ channels the results are different 
for the single top and single antitop cross sections.

In the $t$ channel the threshold approximation is not good at the LHC energy,
and thus we do not include soft-gluon corrections.
The exact NLO cross section for single top is
$\sigma^{t-{\rm channel}}_{\rm top} (14 \, {\rm TeV}, 
m_t=172 \,{\rm GeV})=149 \pm 5 \pm 3 \; 
{\rm pb}=149 \pm 6 \; {\rm pb}$,
while for single antitop it is 
$\sigma^{t-{\rm channel}}_{\rm antitop} (14 \, {\rm TeV}, 
m_t=172 \,{\rm GeV})=91 \pm 3 \pm 2 \;
{\rm pb}=91 \pm 4 \; {\rm pb}$.
\begin{figure}
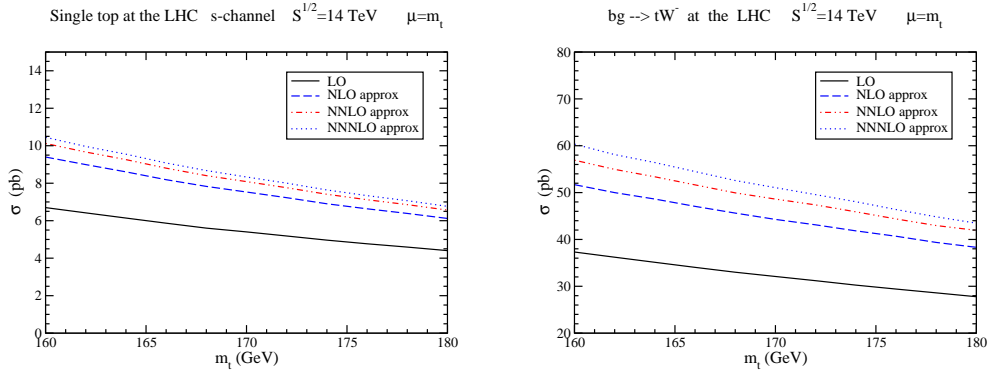

\begin{center}
\includegraphics[width=6cm]{schlhcmtplot.eps}
\hspace{8mm}
\includegraphics[width=6cm]{bglhcmtplot.eps}
\caption{The single top quark cross sections at the LHC 
in the $s$ channel (left) and $tW$ channel (right).}
\end{center}
\end{figure}

In the $s$ channel the cross section, including soft-gluon corrections 
through NNNLO, for single top production is
\beqa
\sigma^{s-{\rm channel}}_{\rm top}(14 \, {\rm TeV}, 
m_t=172\, {\rm GeV})=7.7^{+0.6}_{-0.5} \pm 0.1 \; {\rm pb}
=7.7^{+0.6}_{-0.5} \; {\rm pb} \, ,
\nonumber
\eeqa
while for single antitop production it is 
\beqa
\sigma^{s-{\rm channel}}_{\rm antitop}(14 \, {\rm TeV}, 
m_t=172 \, {\rm GeV})=4.3 \pm 0.1 \pm 0.1 \; {\rm pb}
=4.3 \pm 0.2 \; {\rm pb} \, .
\nonumber 
\eeqa
In the $tW$ channel the cross section, including soft corrections 
through NNNLO, is
\beqa
\sigma^{tW}(14 \, {\rm TeV}, m_t=172 \, {\rm GeV})=43 \pm 5 \pm 1  
\; {\rm pb}=43 \pm 5 \; {\rm pb} \, ,
\nonumber 
\eeqa
with an identical result for ${\bar t} W$ production.

In Figure 5 we plot results for the $s$ channel and $tW$ production 
at the LHC.

\section{Bottom and charm quark pair production}

We now turn to bottom pair and charm pair production at 
fixed-target facilities. In Figure 6 we show results for the theoretical 
cross section \cite{NKRVbc} in $pp$ collisions versus $\sqrt{S}$ using 
the MRST 2002 NNLO pdf \cite{MRST2002}, 
and compare with some experimental data.
For both $b{\bar b}$ and $c {\bar c}$ production there are large 
theoretical and experimental uncertainties.
\begin{figure}
\begin{center}
\includegraphics[width=3.8cm]{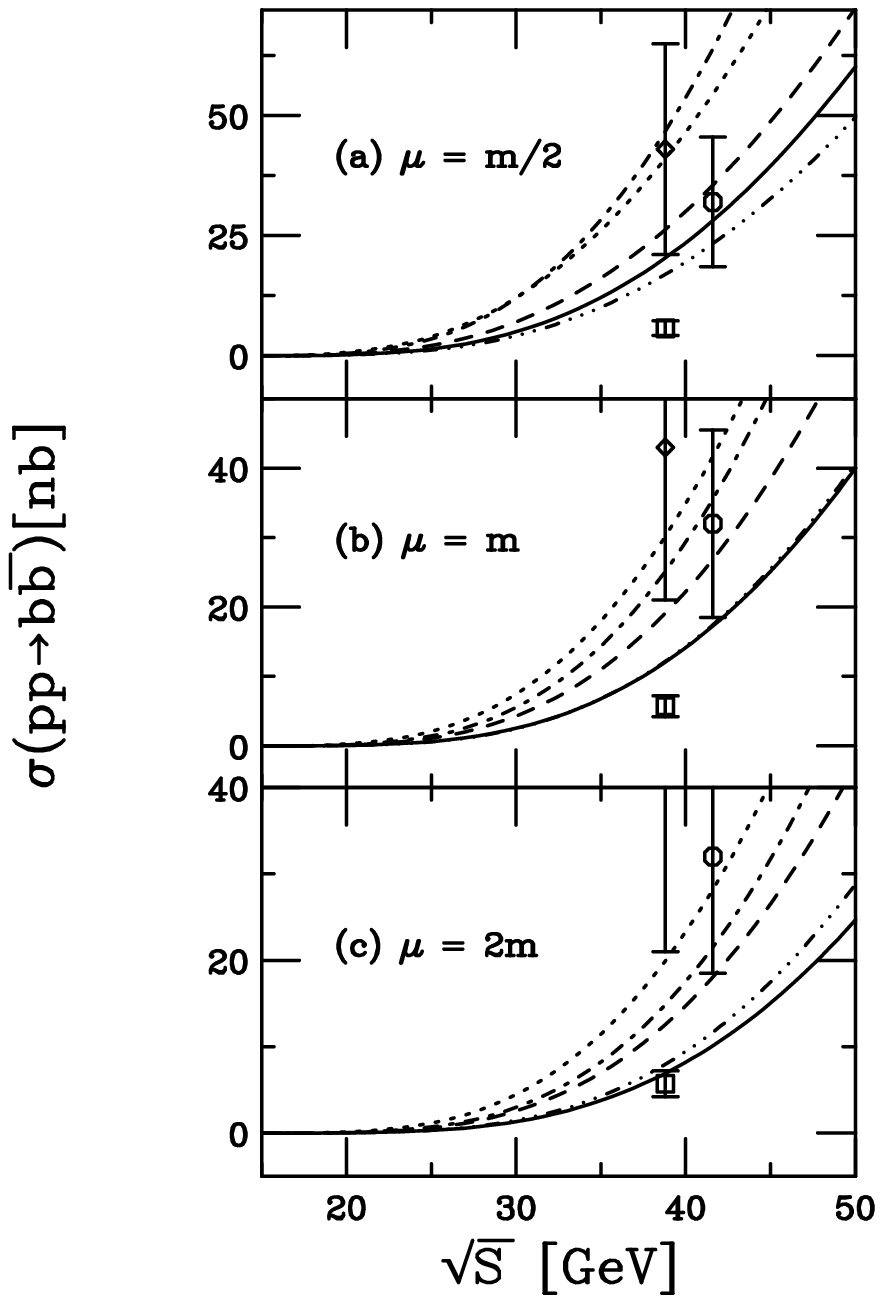}
\hspace{8mm}
\includegraphics[width=7cm]{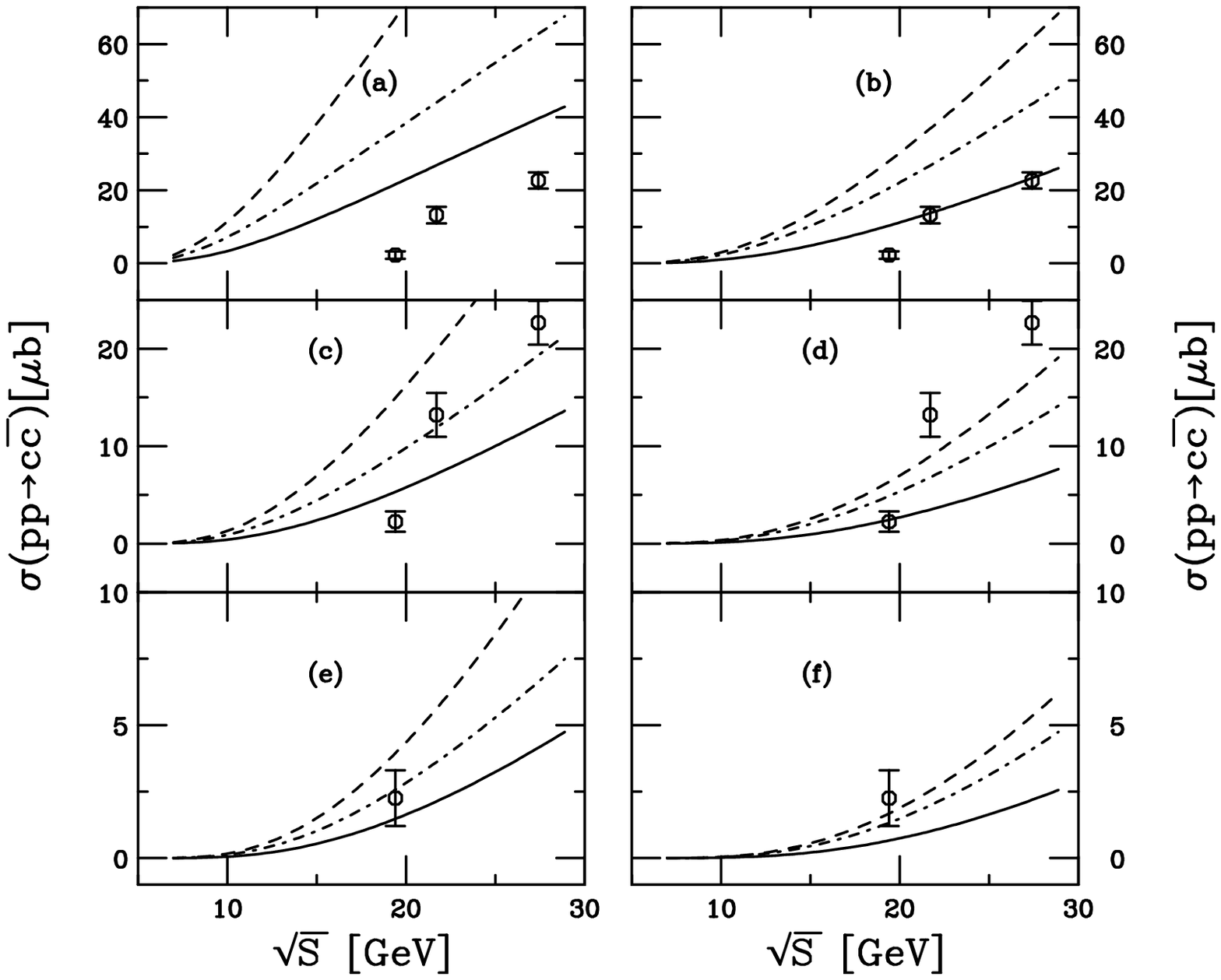}
\caption{(Left) The $b{\bar b}$ cross section in $pp$ collisions. Shown are 
NLO (solid) and NNLO-NNLL (dot-dashed) results for $m_b=4.75$ GeV, 
and NNLO-NNNLL+$\zeta$ results for $m_b=4.5,4.75,5$ GeV (other curves)
with $\mu=m_b/2$ (a), $\mu=m_b$ (b) and $\mu=2 m_b$ (c).  
(Right) The $c{\bar c}$ cross section in $pp$ collisions.  Shown are 
NLO (solid), NNLO-NNLL (dashed) and NNLO-NNNLL+$\zeta$ (dot-dashed)
results for $m_c=1.2$ GeV (a,b), 1.5 GeV (c,d) and 1.8 GeV (e,f), 
with $\mu=m_c$ (a,c,e) and $\mu=2m_c$ (b,d,f).}
\end{center} 
\end{figure}

For the HERA-B experiment the theoretical cross section is \cite{NKRVbc}
\beqa
\sigma^{pp \rightarrow {b \bar b}}(41.6 \, {\rm GeV}, 
m_b=4.75 \, {\rm GeV})=28 \pm 9 {}^{+15}_{-10} \; {\rm nb} \, ,
\nonumber 
\eeqa
where the first uncertainty is due to scale variation and the second 
is due to uncertainty in the bottom quark mass, $4.5 \le m_b \le 5$ GeV.
This is in agreement with HERA-B data \cite{HERAB}.

\section{Two-loop soft-gluon resummation for heavy quarks}

Further progress in the soft-gluon resummation program at NNLL 
requires two-loop calculations in the eikonal approximation.
Resummation is controlled by a soft anomalous dimension matrix, $\Gamma_S$ 
\cite{NKGS}, which can be calculated by the evaluation of 
dimensionally regularized graphs with Wilson lines.
\begin{figure}
\includegraphics[width=7cm]{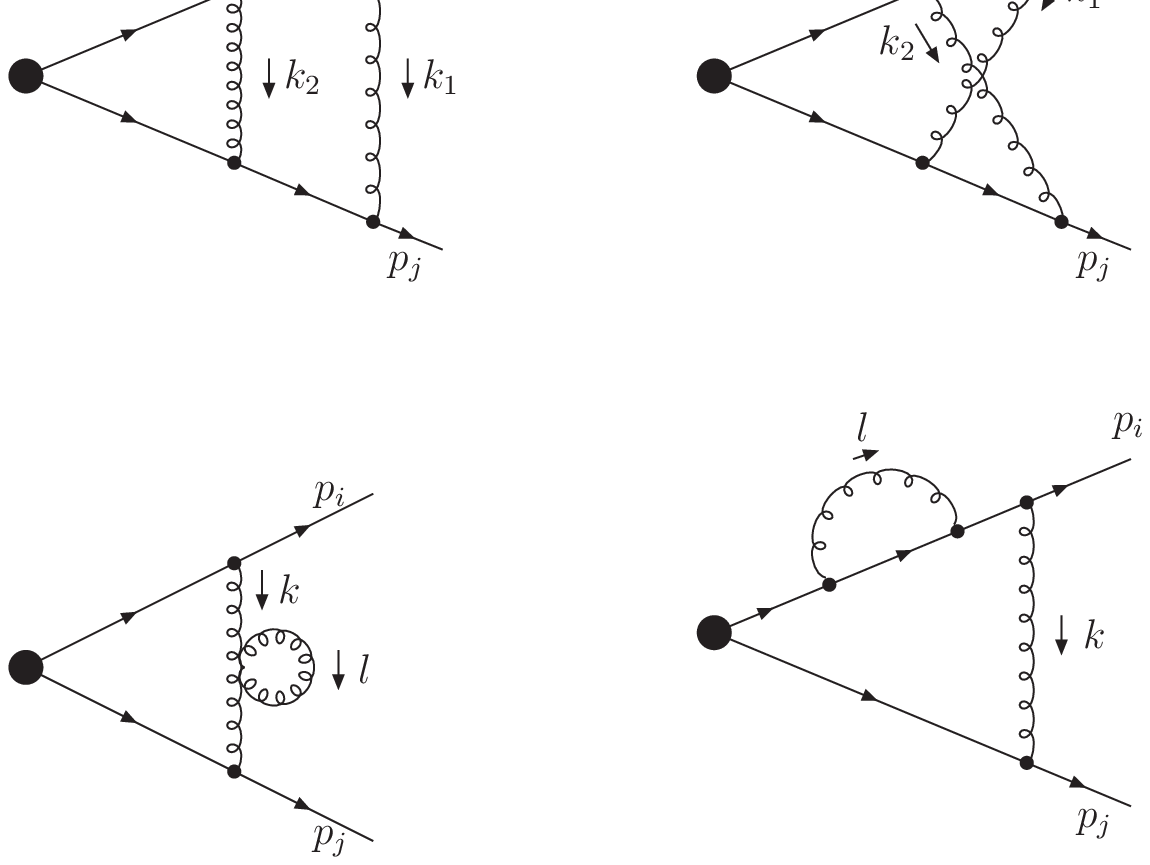}
\hspace{-23mm}
\includegraphics[width=7cm]{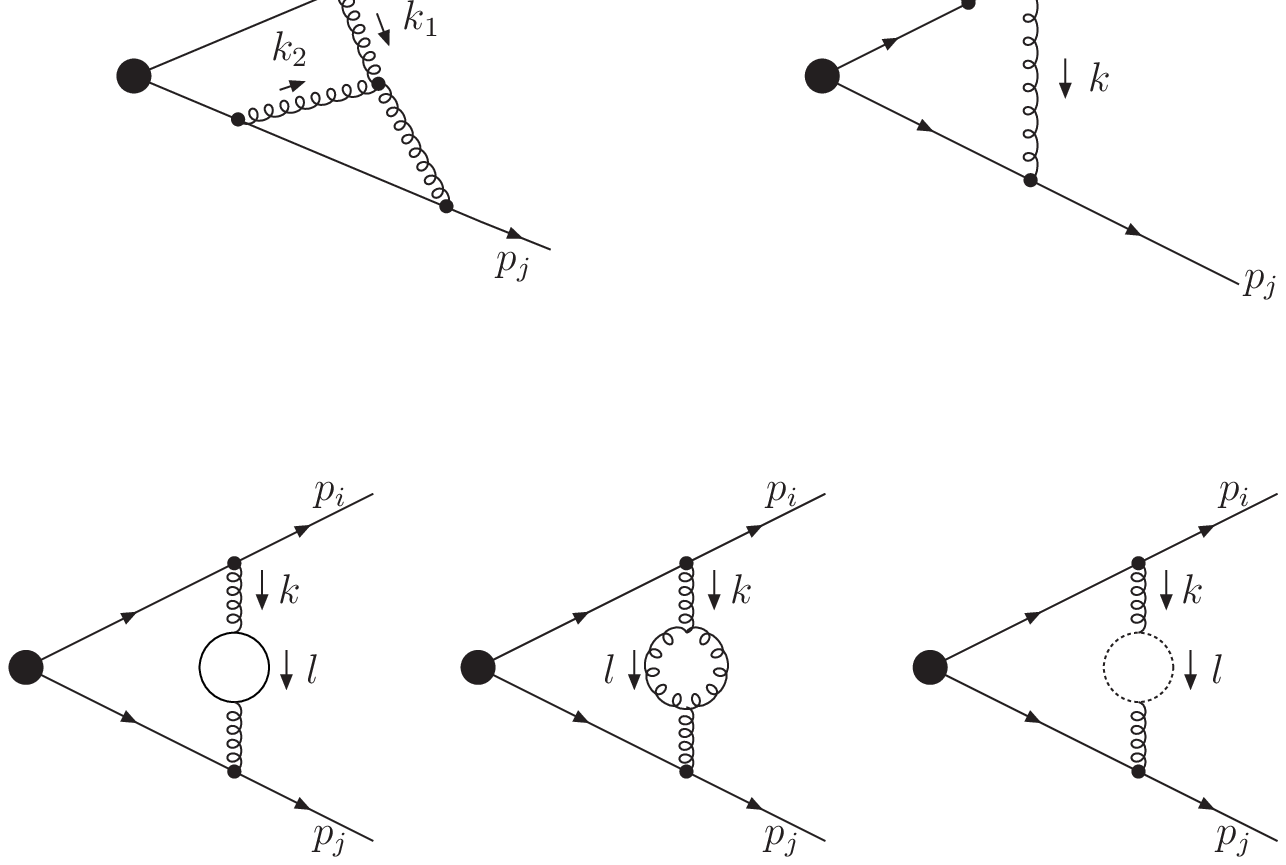}
\vspace{-63mm}
\caption{Two-loop diagrams with heavy-quark eikonal lines.}
\end{figure}

Figure 7 shows some typical two-loop diagrams with heavy-quark lines; 
there are additional graphs not shown here including heavy-quark self-energies 
and one-loop counterterms.
For massless quarks $\Gamma_S$ was calculated at two loops in Ref. \cite{ADS} 
and shown to be proportional to the one-loop result. 
The mass makes the corresponding calculation for heavy quarks more difficult  
but several results for the required integrals have appeared in 
Refs. \cite{KSS,NKPS} and work is ongoing. 
It is clear that the two-loop $C_F n_f$ terms in 
the massive case obey the same relation with respect to the one-loop result 
as in the massless case, but the $C_F C_A$ terms are more challenging.


\vspace{-5mm}

\begin{thebibliography}{00}

\bibitem{NKRV08}
N. Kidonakis and R. Vogt, Phys. Rev. D 78, 074005 (2008), 
arXiv:0805.3844 [hep-ph].

\bibitem{CDF}
CDF Collaboration,  Phys. Rev. Lett. 96, 202002 (2006)
[hep-ex/0603043];
Phys. Rev. D 74, 072006 (2006) [hep-ex/0607035];
D 74, 072005 (2006) [hep-ex/0607095];
D 76, 072009 (2007), arXiv:0706.3790 [hep-ex].

\bibitem{D0}
D0 Collaboration, Phys. Rev. D 74, 112004 (2006)
[hep-ex/0611002];
D 76, 072007 (2007) [hep-ex/0612040];
D 76, 092007 (2007), arXiv:0705.2788 [hep-ex];
D 76, 052006 (2007), arXiv:0706.0458 [hep-ex];
Phys. Rev. Lett. 100, 192004 (2008), arXiv:0803.2779 [hep-ex].

\bibitem{MRST2006}
A.D. Martin, W.J. Stirling, R.S. Thorne, and G. Watt, 
Phys. Lett. B 652, 292 (2007), arXiv:0706.0459 [hep-ph].

\bibitem{CTEQ66}
P.M. Nadolsky {\sl et. al.}, Phys. Rev. D 78, 013004 (2008), 
arXiv:0802.0007 [hep-ph].

\bibitem{NKRV03}
N. Kidonakis and R. Vogt, Phys. Rev. D 68, 114014 (2003) [hep-ph/0308222].

\bibitem{CDFD0st}
D0 Collaboration, Phys. Rev. Lett. 98, 
181802 (2007) [hep-ex/0612052];  
Phys. Rev. D 78, 012005 (2008), arXiv:0803.0739 [hep-ex]; 
CDF Collaboration, Phys. Rev. Lett. 101, 252001 (2008),
arXiv:0809.2581 [hep-ex]. 

\bibitem{NKsttev}
N. Kidonakis, Phys. Rev. D 74, 114012 (2006) [hep-ph/0609287].

\bibitem{NKstlhc}
N. Kidonakis, Phys. Rev. D 75, 071501(R) (2007) [hep-ph/0701080].

\bibitem{NKAPP}
N. Kidonakis, Acta Phys. Polon. B 39, 1593 (2008), arXiv:0802.3381 [hep-ph].

\bibitem{MRST2004}
A.D. Martin, R.G. Roberts, W.J. Stirling, and R.S. Thorne,  
Phys. Lett. B 604, 61 (2004) [hep-ph/0410230]. 

\bibitem{NKRVbc}
N. Kidonakis and R. Vogt, Eur. Phys. J. C 36, 201 (2004) [hep-ph/0401056].

\bibitem{MRST2002}
A.D. Martin, R.G. Roberts, W.J. Stirling, and R.S. Thorne,  
Eur. Phys. J. C 28, 455 (2003) [hep-ph/0211080].

\bibitem{HERAB}
HERA-B Collaboration, Phys. Rev. D 73, 052005 (2006) [hep-ex/0512030];
Phys. Lett. B 650, 103 (2007) [hep-ex/0612024].

\bibitem{NKGS}
N. Kidonakis and G. Sterman, Nucl. Phys. B 505, 321 (1997) [hep-ph/9705234]. 

\bibitem{ADS}
S.M. Aybat, L.J. Dixon, and G. Sterman, Phys. Rev. D 74, 074004 (2006)
[hep-ph/0607309].

\bibitem{KSS}
N. Kidonakis, A. Sabio Vera, and P. Stephens, arXiv:0802.4240 [hep-ph].

\bibitem{NKPS}
N. Kidonakis and P. Stephens,  in {\sl DIS 2008}, arXiv:0805.1193 [hep-ph], 
and in preparation. 

\end{thebibliography}
\end{document}